\documentclass[twocolumn,a4paper,superscriptaddress,showkeys]{revtex4}

\usepackage{mathbbol,amssymb,latexsym,amsfonts,amsmath,amsthm}
\usepackage{mathpazo}
\usepackage{graphicx}

\newcommand{\ie}{\textit{i.e.}}

\newcommand{\ecoli}{\textit{E.~coli} }

\newcommand{\R}{\mathbb{R}}
\newcommand{\A}{\mathcal{A}}

\renewcommand{\R}{\mathcal{R}}

\begin{document}

\title{Reverse-engineering transcriptional modules from gene
  expression data}

\date{\today}

\author{Tom Michoel}

\email[Corresponding author, E-mail: ]{tom.michoel@psb.ugent.be}

\affiliation{Department of Plant Systems Biology, VIB, Technologiepark
  927, B-9052 Gent, Belgium}

\affiliation{Department of Molecular Genetics, UGent, Technologiepark
  927, B-9052 Gent, Belgium}

\author{Riet De Smet}

\affiliation{CMPG, Department Microbial and Molecular Systems,
  KULeuven, Kasteelpark Arenberg 20, B-3001 Leuven, Belgium}

\author{Anagha Joshi}

\affiliation{Department of Plant Systems Biology, VIB, Technologiepark
  927, B-9052 Gent, Belgium} 

\affiliation{Department of Molecular Genetics, UGent, Technologiepark
  927, B-9052 Gent, Belgium}

\author{Kathleen Marchal}

\affiliation{CMPG, Department Microbial and Molecular Systems,
  KULeuven, Kasteelpark Arenberg 20, B-3001 Leuven, Belgium}

\affiliation{ESAT-SCD, KULeuven, Kasteelpark Arenberg 10, B-3001
  Leuven, Belgium}

\author{Yves Van de Peer}

\affiliation{Department of Plant Systems Biology, VIB, Technologiepark
  927, B-9052 Gent, Belgium}

\affiliation{Department of Molecular Genetics, UGent, Technologiepark
  927, B-9052 Gent, Belgium}

\begin{abstract}
  ``Module networks'' are a framework to learn gene regulatory
  networks from expression data using a probabilistic model in which
  coregulated genes share the same parameters and conditional
  distributions. We present a method to infer ensembles of such
  networks and an averaging procedure to extract the statistically
  most significant modules and their regulators. We show that the
  inferred probabilistic models extend beyond the data set used to
  learn the models.
\end{abstract}


\maketitle

\section{Introduction}
\label{sec:intro}

Methods for reverse-engineering transcription regulatory networks from
high-throughput microarray data come in many different flavors
\cite{friedman2004,gardner2005,lemm06,bansal2007,bussemaker2007}.  An
important class of methods are those which not only seek to identify
the topological wiring of the network
\cite{butte2000,basso2005,faith2007}, but also attempt to infer a
model of the biological system which explains the observed gene
expression patterns and generates testable hypotheses.  Such models
can take the form of probabilistic graphical models
\cite{segal2003,friedman2004,beer2004}, simplified kinetic equation
models \cite{bonneau2006}, or biophysical models
\cite{bussemaker2007}. A common property of all modeling approaches is
that the number of parameters is much larger than the number of
experimental data points available to define them.  Dimensionality
reduction is usually achieved by a coarse-graining step which
collapses individual genes into clusters of coexpressed genes or
modules, where all genes in a cluster share the same model parameters
\cite{vandenbulcke2006}.  This conceptual simplification has as a
drawback that inferred interactions are influenced by the module
quality. Moreover, it is hard to translate the concept of a biological
module in a strict mathematical definition. When searching for
modules, often many local optima exist with partially overlapping
modules differing from each other in a few genes or conditions.
Therefore, in our approach we exploit the ``fuzzy'' property of a
module to increase the reliability of the predicted interactions.
Instead of reporting only one cluster solution (local optimum), we use
a stochastic approach to generate many partially redundant cluster
solutions (bootstrapping) and generate an ensemble solution by
averaging over multiple high-scoring models.  Crucial for the success
of the ensemble approach is the availability of an efficient method
for sampling a large number of different models covering the whole
search space of possible models \cite{dietterich1997}. Therefore we
use the deterministic approach of \cite{michoel2007a} as a basis for
our sampling method.  In this approach gene and condition clustering
are decoupled from learning the regulatory programs compared to the
original method of \cite{segal2003} which optimizes both
simultaneously.  This allows for a higher efficiency while maintaining
an equal performance rate \cite{michoel2007a}.  Several extensions of
the method of \cite{segal2003} infer transcriptional modules from gene
expression and genome wide location analyis
\cite{barjoseph2003,xu2004,gao2004,wu2006}. It is entirely feasible to
develop ensemble strategies for these methods as we did for
\cite{segal2003}. This constitutes an interesting line of future
research. In this paper we give a brief description of the algorithm
and highlight some results for an expression compendium for \textit{E.
  coli} \cite{faith2007}. A complete analysis of this data set and
comparison of our approach to the mutual information based CLR method
\cite{faith2007} is given in \cite{michoel2008}. A detailed comparison
between the ensemble approach and the direct-optimization based method
of \cite{segal2003} on \textit{S. cerevisiae} data is given in
\cite{joshi2008}.

\section{Results and discussion}
\label{sec:results}

\subsection{The algorithm}
\label{sec:lemone-algorithm}

The algorithm takes as input a gene expression data set and a list of
candidate regulators. It gives as output a large number of
probabilistic models consisting of a set of gene clusters, with for
each gene cluster a partition of the experiments and a probabilistic
regulatory program explaining the observed experiment partitions in
terms of the expression of a small number of regulators. The number of
gene and experiment clusters is determined automatically and can vary
from one solution to the next.  Using overrepresentation in the
ensemble, the most probable interactions can be identified.  The
regulatory programs can be validated on new experimental data and
generate testable hypotheses about conditional regulation of the
inferred gene clusters.

\begin{figure}[ht!]
  \centering
  \includegraphics[width=\linewidth]{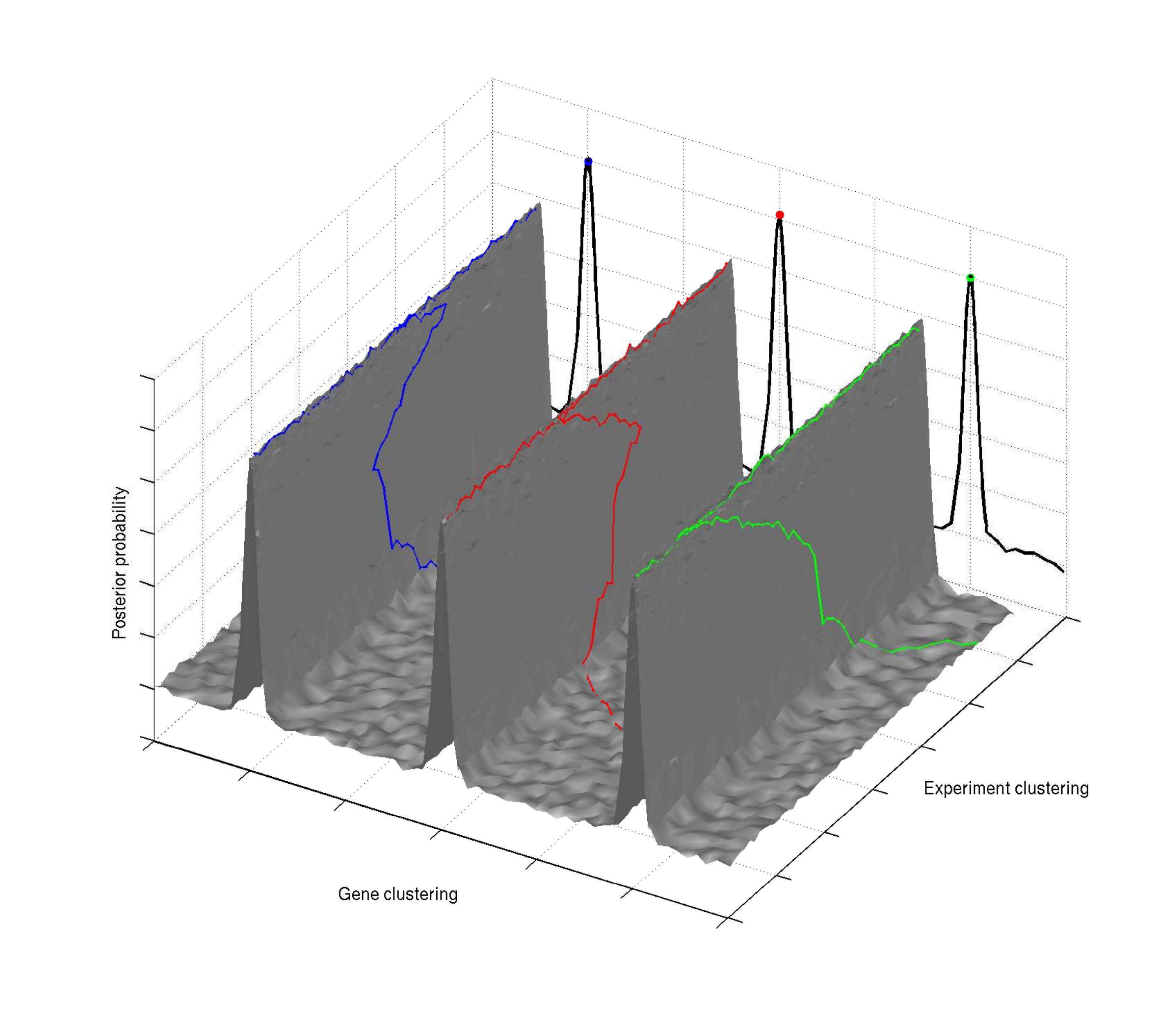}
  \caption{Graphical cartoon representation of the Gibbs sampling
    procedure for two-way clustering of genes and conditions
    \cite{joshi2007}. Each colored curve represent a random restart
    converging on a distinct local optimum in the direction of genes.
    In the direction of experiments the whole search space can be
    covered in one run.}
  \label{fig:gibbs}
\end{figure}

The first step of the algorithm consists of generating an ensemble of
gene clusters with experiment partitions.  A Gibbs sampling method
iteratively updates the assignment of each gene given the current gene
and experiment clusters, and the assignment of each experiment in each
gene cluster given the current assignment of all other experiments in
that gene cluster, iterating until a stationary state is reached.
Details about the Gibbs sampler algorithm and a complete analysis of
its convergence properties can be found in \cite{joshi2007}. Briefly,
in a single run, the Gibbs sampler reaches a local optimum in the
direction of genes, but covers the whole search space in the direction
of experiments. This implies that for a given cluster of coexpressed
genes there are multiple equiprobable ways of partitioning the
experiments. To also sample from the whole search space in the
direction of genes, we perform several independent Gibbs sampler runs
with random restarts. In \cite{joshi2007}, it is shown that each of
the local optima in the direction of genes is (approximately) of the
same height, and therefore equally important, and that a relatively
small number of local optima is sufficient to cover the whole search
space (typically two distinct sets of 10 local optima for a data set
of 1000 genes agree for 95\% on the probability for each pair of genes
to be clustered together, see \cite{joshi2007} for details).  A
graphical cartoon representation of the Gibbs sampling procedure is
given in Figure \ref{fig:gibbs}.

In the second step of the algorithm, regulatory programs are learned
for each experiment partition for each gene cluster. This is achieved
by linking the sets in the experiment partition hierarchically in a
decision tree. For each split in this tree a candidate regulator is
found whose expression is significantly different on both sides of the
split, as measured by an entropy measure.  Details can be found in
\cite{michoel2007a}.

\begin{figure}[ht!]
  \centering
  \includegraphics[width=\linewidth]{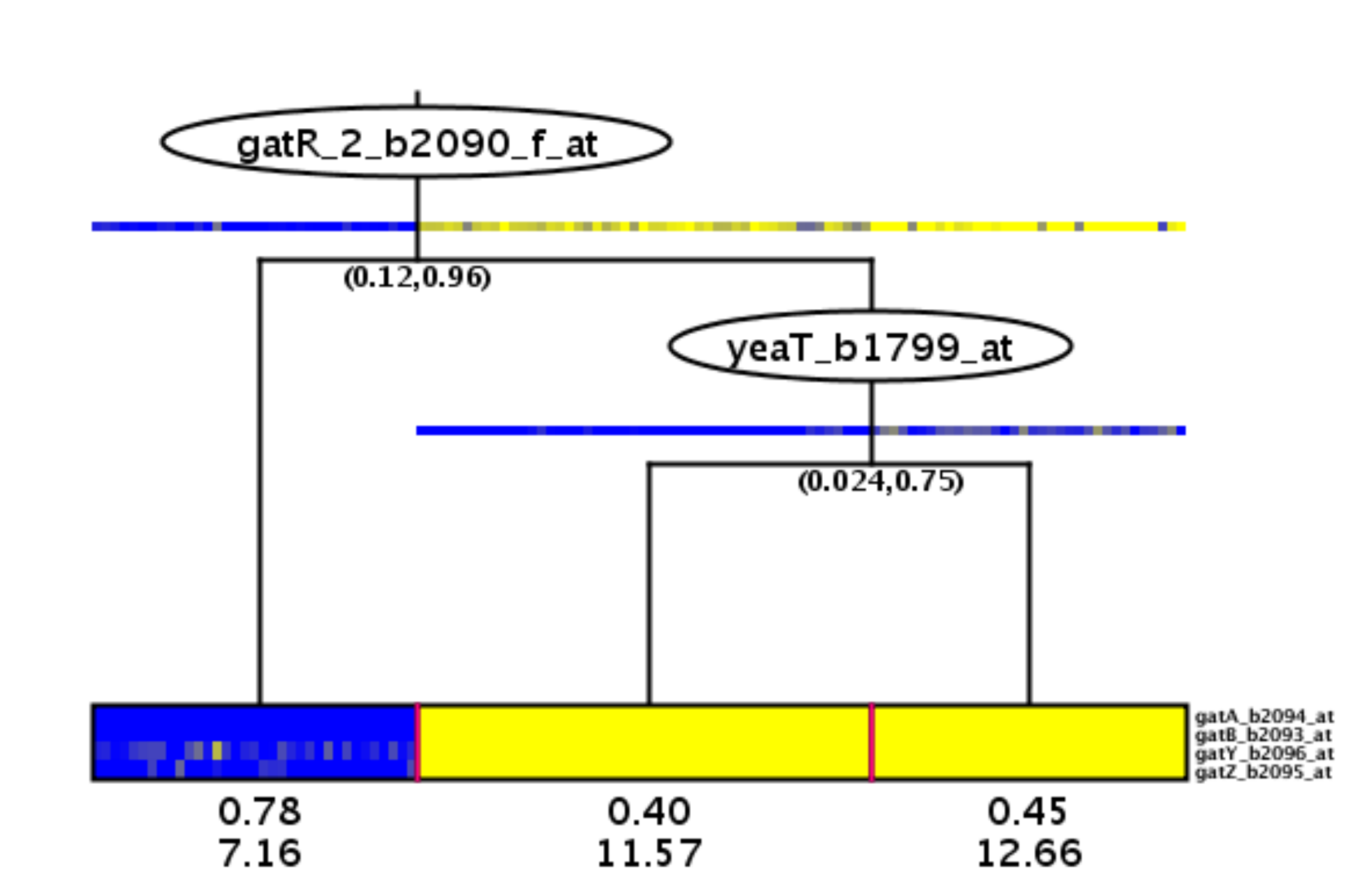}
  \caption{Example module with a regulatory program. The numbers under
    the experiment clusters are the standard deviation and mean
    expression value of the data in the cluster. The numbers under
    each tree node are the normalized Bayesian score gained in the
    Gibbs sampler by making this data split \cite{joshi2007}, and a
    percentage quality score for the assignment of this regulator
    \cite{michoel2007a}. The colored bars on each tree node are the
    expression levels of the regulators with experiments sorted in the
    same order as in the experiment clusters.}
  \label{fig:module}
\end{figure}

A gene cluster with a regulatory program is called a
\emph{transcriptional module} and a partition of all genes into
clusters, each with a regulatory program, a \emph{module network}.  A
sample module is shown in Figure \ref{fig:module}.  To each module
network corresponds a probabilistic model defined by a probability
density function $p(x_1,\dots,x_N)$ with $x_i$ the continuous valued
expression level of gene $i$ (see Methods). The value of
$p(x_1,\dots,x_N)$ measures how well the model explains a particular
experiment with expression levels for all genes.

\subsection{Ensemble averaging}

We have applied the algorithm on a compendium of Affymetrix
microarrays for \ecoli \cite{faith2007}. The compendium contains 445
expression profiles for 4345 genes under 189 different stress
conditions and genetic perturbations. The results are validated by
comparison with existing knowledge of transcriptional interactions in
RegulonDB \cite{salgado2006} and EcoCyc \cite{keseler2005}. A complete
analysis of the inferred module networks and their biological
significance is beyond the scope of this paper and will be given
elsewhere \cite{michoel2008}. Here we highlight a few examples of
inferred interactions which are representative for the working
mechanism of the ensemble approach.

\begin{figure}[ht!]
  \centering
  \includegraphics[width=\linewidth]{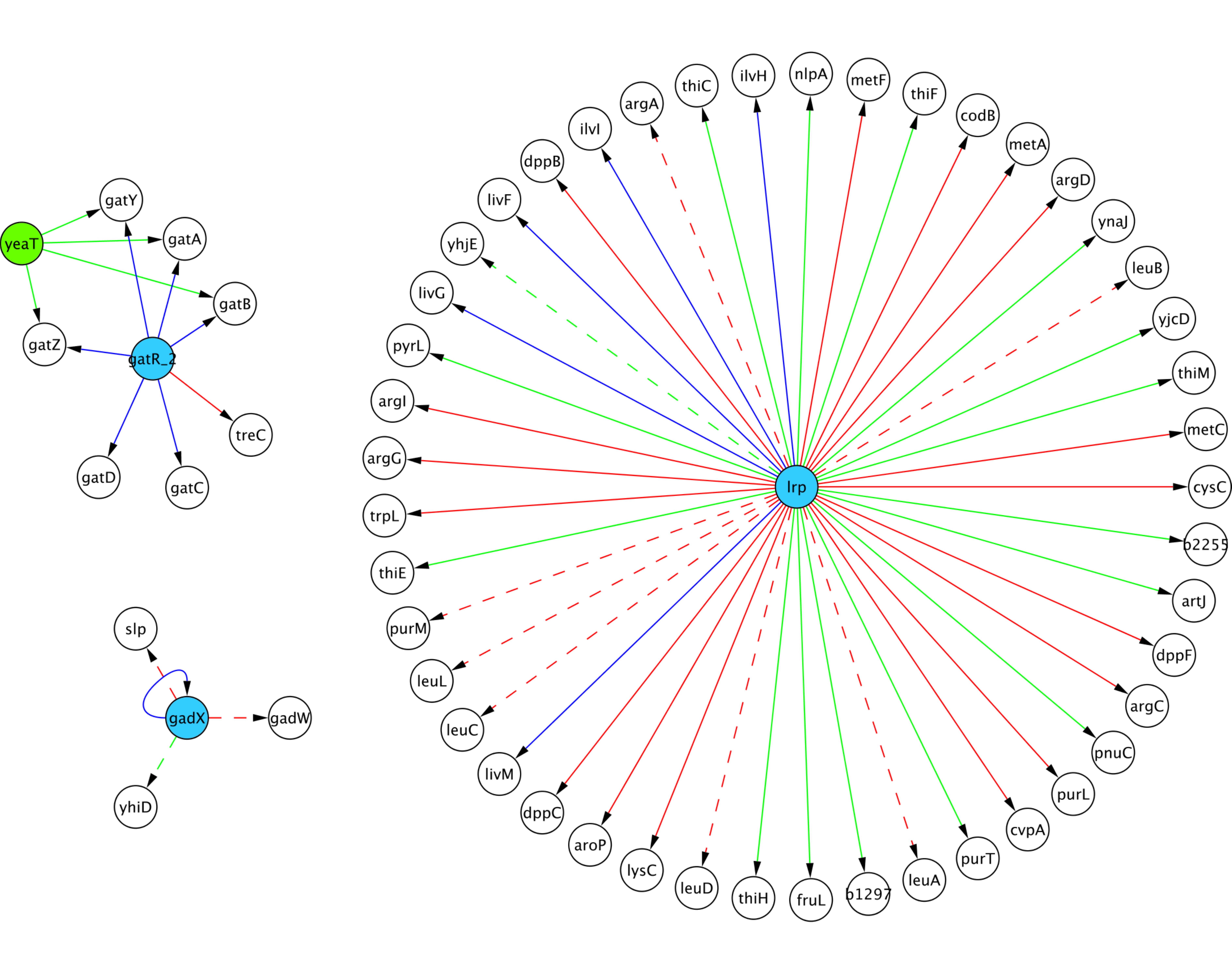}
  \caption{Sample interactions discussed in the main text. Blue edges
    are present in RegulondDB, green edges connect to genes not in
    RegulonDB, and dashed edges are new predictions validated by
    literature.}
  \label{fig:gatR_gadX_lrp}
\end{figure}

A first example is given by the \emph{gatYZABCD} operon (involved in
galactitol metabolism). As expected the genes in this operon
consistently cluster together. GatR\_2, a subunit of the GatR
transcriptional repressor, is consistently assigned as a candidate
regulator. It is known that GatR negatively controls transcription
initiation of the \emph{gatYZABCD} operon, which is the only known
target of GatR according to RegulonDB.  The many local optima,
generated by the Gibbs sampling approach resulted in 81 additional
genes which were clustered at least once with one of the
\emph{gatYZABCD} genes. Except for one potential new target
\emph{treC} which was found repeatedly in combination with \emph{gatC}
and \emph{gatD}, the interactions between the remaining 80 genes and
GatR were not sufficiently significant to be retained in the final
solution. This illustrates how the ensemble approach guarantees an
effective filtering of false positives while retaining the true
positives.  This example also illustrates how combinatorial regulation
can be detected. For \emph{gatY}, \emph{gatZ}, \emph{gatA} and
\emph{gatB}, a second regulator, YeaT (a predicted regulator involved
in malate metabolism), with an anticorrelated expression pattern
seemed to play a role in a subset of the conditions (see Figure
\ref{fig:module}).

Four interactions could be inferred for GadX (a transcriptional
activator involved in acid resistance) : GadX itself, which makes
sense as it is known to be autoregulated \cite{tramonti2002}, and
three novel targets \emph{slp}, \emph{gadW} and \emph{yhiD}. This
finding is supported by literature, as the expression of both
\emph{slp} \cite{tucker2003} and \emph{yhiD} \cite{tucker2003} seemed
to be affected in a \emph{gadX} mutant, while \emph{gadX} and
\emph{gadW} seem to tightly control each other’s expression
\cite{foster2004}.

For Lrp (a regulator involved in the high-affinity transport of
branched-chain amino acids and a mediator of the leucine response) 44
interactions could be found of which 5 could be confirmed by RegulonDB
(\emph{ilvI}, \emph{ilvH}, \emph{livG}, \emph{livM} and \emph{livN})
and 39 were new. Among the predicted interactions with Lrp we found
the \emph{leuLABCD} genes.  According to literature, the Lrp dependent
regulation of the \emph{leuLABCD} operon is only indirect
\cite{landgraf1999}. However, without additional data, no distinction
can be made between direct and indirect effects of a regulator if both
give rise to a correlated expression level with the targets. Four of
our predicted targets (\emph{purM}, \emph{argA}, \emph{yhjE} and
\emph{aroP}) were tested by ChIP analysis \cite{faith2007} and three
(\emph{purM}, \emph{argA} and \emph{yhjE}) were confirmed.
\emph{YhjE} was also found to be differentially expressed in a
microarray analysis of \emph{lrp} mutants \cite{hung2002}. For the
remaining genes no clear indication for their regulation by Lrp could
be found. However, it should be noted that Lrp not only acts as a
regulator, binding specific DNA-sequences, but also functions as a
DNA-organizing protein, extending its global role in regulation
\cite{peterson2007,pul2007}.  A large regulon of Lrp, as was detected
by our method, thus is in line with this more global role of Lrp.

\subsection{Model evaluation}
\label{sec:model-eval}

\begin{figure}[ht!]
  \centering
  \includegraphics[width=\linewidth]{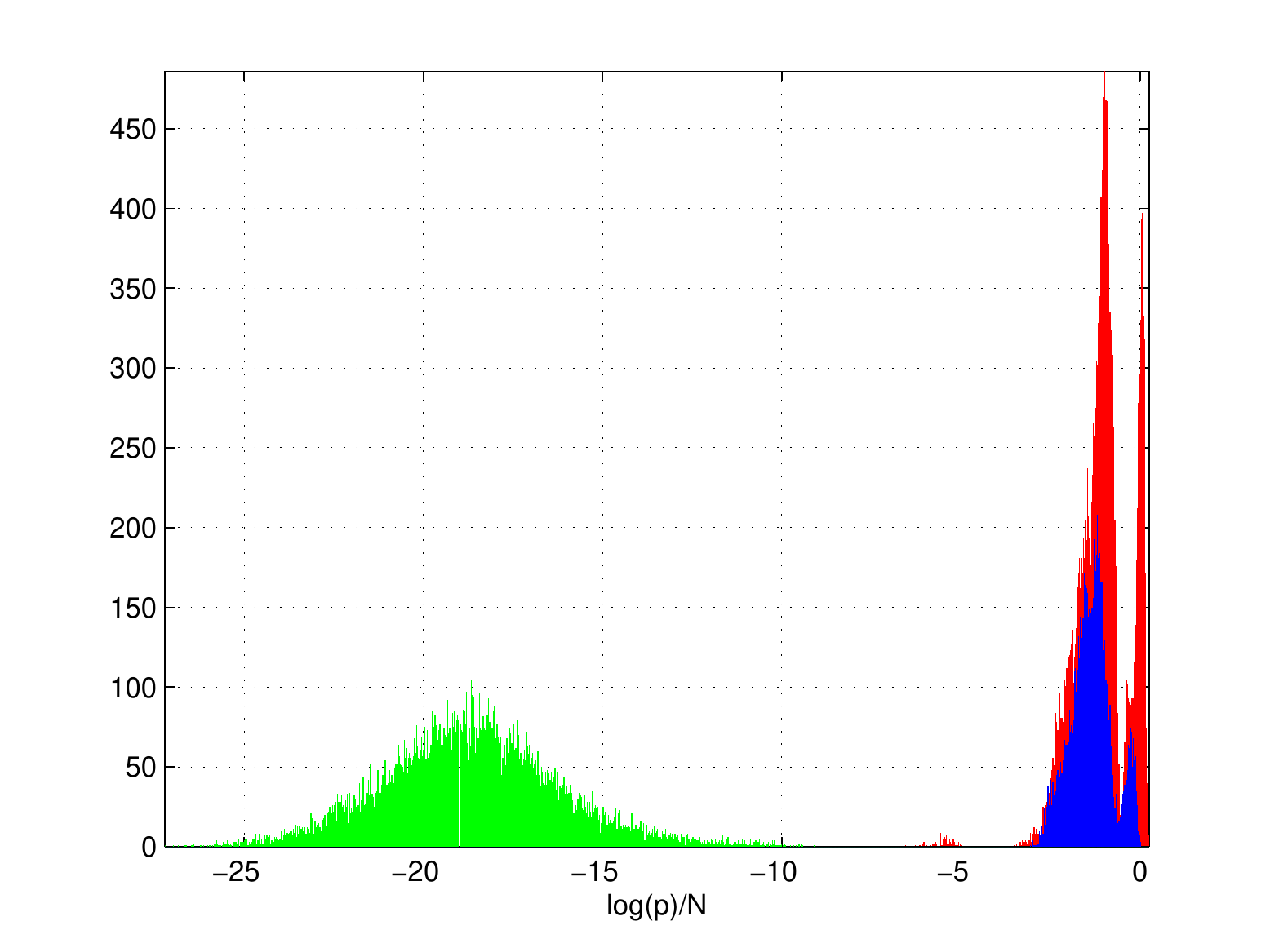}
  \caption{Histogram of $\frac1N\log p(x_1,\dots,x_N)$ for 100 models
    in the ensemble, for the original data (red), new data (blue), and
    randomized data (green, around 15\% zero likelihood values ($\log
    p=-\infty$) not shown).}
  \label{fig:eval}
\end{figure}

One of the purposes of model-based reverse-engineering methods is to
infer a model of the system which extends beyond the data set used to
learn the model. As such they can form the basis for developing
methods which use new data to refine and extend a partially validated
model, rather than infer a completely new network model each time a
data set is updated.  Validation of a model is done by comparing the
distribution of $\frac1N\log p(x_1,\dots,x_N)$ (see Methods) for new
data with the distribution for data used to learn the model. In
general, higher values of $\log p$ means better explanation of the
data by the model. The result of this comparison for 75 experiments
recently added to the $M^{3D}$ database \cite{faith2007} with the 189
original experiments, for 100 module network models selected at random
from the ensemble, is shown in Figure \ref{fig:eval}.  The overlap
between the distributions shows that the probabilistic models indeed
generalize to unseen data.  For comparison we also perform a
randomization test by permuting the gene indices of the data matrix.
As expected, the evaluation of the models on randomized data is
several orders of magnitude smaller than on real data, and in fact
around 15\% of the randomized experiments have a value $\log
p=-\infty$, \ie, zero likelihood.

\section{Methods}
\label{sec:methods}

\subsection{Data sets}

We downloaded a compendium of expression profiles for \ecoli
\cite{faith2007} and a list of $328$ candidate regulators from
\url{http://gardnerlab.bu.edu/netinfer_plos_2007/}. We averaged over
replicate experiments to obtain a data matrix of 4345 genes and 189
experiments.  Additional data for 4292 genes and 75 experiments for
validation of the probabilistic models was downloaded from
\url{http://m3d.bu.edu/norm/}.

\subsection{Probabilistic model evaluation}

The probabilistic model introduced in \cite{segal2003} associates to
each gene $i$ a continuous valued random variable $X_i$ measuring the
gene's expression level. The distribution of $X_i$ depends on the
expression level of a set of parent genes chosen from a list of
candidate regulators. Genes regulated by the same parents form a
cluster and share the same model parameters. The joint probability
distribution for the expression levels of all genes decomposes as a
product of conditional distributions,
\begin{equation}\label{eq:2}
  p(x_1,\dots,x_N) = \prod_{k=1}^K \prod_{i\in\A_k} p_k 
  \bigl(x_i \mid \{x_r\colon r\in\R_k\}\bigr)\;,
\end{equation}
where $\{\A_k,k=1,\dots,K\}$ is the set of clusters (\ie, a partition
of the gene set $\{1,\dots,N\}$), and $\R_k$ is the set of regulators
for cluster $k$. The distribution \eqref{eq:2} is normalized if the
network from parents to children is acyclic.  The conditional
distribution $p_k$ of the expression level of the genes in cluster $k$
is a normal distribution with parameters determined by the expression
levels of the parents $\R_k$:
\begin{equation}\label{eq:4}
  p_k\bigl(x \mid \{x_r\colon r\in\R_k\}\bigr)= p(x\mid \mu_\ell,
  \tau_\ell)\;.
\end{equation}
The parameters $\mu_\ell$ and $\tau_\ell$ are determined by arranging
the parents in a decision tree. The tests on the nodes of the decision
tree are of the form `$x_r>z$ ?' for some threshold value $z$, where
$x_r$ is the expression value of the parent $r$ associated to the
node. The leaves $\ell$ of the decision tree are the sets of an
experiment partition for cluster $k$ and $\mu_\ell$ and $\tau_\ell$
are the mean, resp. precision of the expression levels of the genes in
the cluster in this subset of experiments. See \cite{michoel2007a} for
more details.

To evaluate a model of the form (\ref{eq:2}), we are only interested
in genes for which the model makes actual predicitions, \ie, genes
belonging to clusters with a regulation tree. If the clustering
procedure does not find distinct experiment clusters for a certain
gene cluster, the model predicts one broad normal distribution for the
genes in this cluster. Any expression data for these genes will fit
the model and thereby obscure the signal of the genes for which true
predictions are made.  For the particular data used in Section
\ref{sec:model-eval}, the number of genes in the new data set is 4292
vs. 4345 in the data used to learn the models.  Six of the missing
genes belong to the regulator list for the learned models.  Hence in
some models we may not be able to compute the conditional probability
distributions for all genes. Altogether around 2700 genes remain for
each model.

\subsection{Software availability}

The Java software package LeMoNe for learning module networks is
available from our website
\url{http://bioinformatics.psb.ugent.be/software/details/LeMoNe}.

\begin{acknowledgments}
  RDS is a research assistant of the IWT. AJ is supported by an
  Early-Stage Marie Curie Fellowship.  This work is supported by 1)
  Research Council KUL: GOA AMBioRICS,GOA/08/011, CoE EF/05/007
  SymBioSys, 2) FWO: projects G.0318.05, 3) IWT: SBO-BioFrame, 4) IUAP
  P6/25 (BioMaGNet).
\end{acknowledgments}


\end{document}